\documentclass[12pt]{iopart}
\usepackage{lscape}
\usepackage{iopams}
\usepackage{graphicx}
\RequirePackage{xspace}
\newcommand{\ket}[1]{\ensuremath{|#1\rangle}\xspace}

\begin{document}
\title[Exploring Biorthonormal Transformations of \mbox{Pair-Correlation} Functions \ldots]{Exploring Biorthonormal Transformations of \mbox{Pair-Correlation} Functions in Atomic Structure Variational Calculations}
\author{S. Verdebout$^{1}$, P. J\"onsson$^{2}$, G. Gaigalas$^{3}$, M. Godefroid$^{1}$  and C. Froese Fischer$^{4}$ }

\address{$^{1}$ Chimie Quantique et Photophysique, CP160/09, Universit\'e Libre de Bruxelles, \\ 
Av. F.D. Roosevelt 50, B-1050 Brussels, Belgium} 

\address{$^{2}$  Nature, Environment, Society, Malm\"{o} University, 205-06 Malm\"{o}, Sweden }

\address{$^{3}$ Vilnius University Research Institute of Theoretical Physics and Astronomy, \\ 
A. Go\v{s}tauto 12, LT-01108 Vilnius, Lithuania}

\address{$^{4}$ National Institute of Standards and Technology \\
           Gaithersburg, MD 20899-8420, USA }

\ead{mrgodef@ulb.ac.be}
\begin{abstract}
Multiconfiguration expansions frequently target valence correlation and correlation between valence electrons and the outermost core electrons. Correlation within the core is often neglected.  A large orbital basis is needed to saturate both the valence and core-valence correlation effects. This in turn leads to huge numbers of CSFs, many of which are unimportant. To avoid the problems inherent to the use of a single common orthonormal orbital basis  for all correlation effects in the MCHF method, we propose to optimize independent MCHF pair-correlation functions (PCFs), bringing their own orthonormal one-electron basis. Each PCF is generated by allowing single- and double- excitations from a multireference (MR) function.  This computational scheme has the advantage of using targeted and optimally localized orbital sets for each PCF. These pair-correlation functions are 
coupled together and with each component of the MR space through a low dimension generalized eigenvalue problem.  Nonorthogonal orbital sets being involved, the interaction and overlap matrices are built using biorthonormal transformation of the coupled basis sets followed by a counter-transformation of the PCF expansions. 

Applied to the ground state of beryllium, the new method gives total energies that are
lower than the ones from traditional CAS-MCHF calculations using large  orbital active sets. It is fair to say that we now have the possibility to account for, in a balanced way, correlation deep down in the atomic core in variational calculations.
\end{abstract}
\pacs{31.15.ac, 31.15.V-, 31.15.xt}

\submitto{\jpb}
\noindent{\it Keywords\/}:
Electron correlation, pair-correlation function, biorthonormal transformation.

\vfill
\noindent{\hfill \bf \today}

\maketitle


\section{Introduction}

There are several widely used methods
for calculating many-electron atoms, but many-body effects still represent a real challenge.
The many-body perturbation theory (MBPT) appears to
be the most effective for atoms with one valence electron but its
accuracy  is not satisfactory for the atoms with more than
one valence electron, mainly because of the poor convergence of the MBPT for
the valence-valence correlations~\cite{Koz:04a}.
However, the core-valence correlations still can be
effectively treated with MBPT. For this reason it
was suggested to combine many-body perturbation theory 
for the core-valence correlations with the configuration interaction (CI) for valence-valence correlations within the
(CI + MBPT) method \cite{Dzuetal:96a}, but this extension to atoms with more than three electrons in open shells met some difficulties. A method based on the so-called $V^{N-M}$ approximation has been proposed by Dzuba and Flambaum \cite{DzuFla:07a} including core-valence correlations by means of MBPT. 
The most recent extension of MBPT is the development of a configuration-interaction plus all-order method for atomic calculations \cite{Safetal:09a}. This is a theoretical method combining the all-order approach currently used in precision calculations of properties of monovalent atoms with the configuration interaction approach that is applicable for many-electron systems.

The coupled-cluster (CC) approach \cite{LinMor:85a} is an interesting alternative, summing up all orders and taking into account pair correlations. Interesting computational developments have been proposed~\cite{LinMuk:87a,LiPal:03a}, not only in physics but also in quantum chemistry (for a complete review, see \cite{BarMus:07a}). Successful applications do exist~\cite{Majetal:04a,Natetal:07a,SahDas:08a}, with some limitations  for atoms with more complicated electron structures than alkali atoms. According to \cite{Dzuetal:96a}, the most obvious shortcoming of the CC method is the neglect of three-particle correlations. Moreover, it treats the valence-valence and core-valence correlations at the same level of approximation while the
former is much stronger than the latter.

In quantum chemistry, variational complete active space self-consistent field (CASSCF) methods are quite successful for describing small and medium-size molecules but are not sufficient when external (dynamic) correlation must be included~\cite{Roo:87a}. The latter are treated through second order perturbation theory using a single or multireference  state~\cite{Finetal:98a,CelWer:00a}. 
Nonperturbative variational methods  treat many-body effects in an accurate way for valence electrons, in both nonrelativistic and relativistic schemes. In this line, multiconfiguration Hartree-Fock (MCHF) and multiconfiguration Dirac-Hartree-Fock (MCDHF), often combined with CI methods \cite{Froetal:97b,Gra:07a} have been widely used for accurate calculations of many-electron atomic properties, focusing on valence and core-valence correlation of large atomic systems.  The accuracy of CI is limited by the incompleteness of the set of configurations used, if the one-electron orbital basis is complete. A practical limitation is the number of possible configuration state functions (CSFs) that becomes so large for a many-electron atom that one has to select only a small fraction of them~\cite{Fro:93a}. This is often done by neglecting core excitations or only including a limited number of them \cite{GodFro:99a,GodFro:99b}. The CI method is by definition (orbital) basis-dependent, while the multiconfiguration methods are not, the CI problem being iteratively coupled to the orbital optimization~\cite{Fro:77a,Ind:95a}.
As discussed in section~\ref{sec_spatial}, the shape of the resulting orbitals strongly depends on the type of correlation introduced through the multiconfiguration 
expansion~\cite{Godetal:98a}. This is due to the properties of the variational principle applied for deriving the MCHF/MCDHF equations to be solved, with the consequence that a set of radial distributions resulting from a given correlation model/expansion could become inadequate, or at least incomplete, for another model  or for an extension of the original one. 
Other problems encountered in variational multiconfiguration calculations are discussed in the present work, illustrating the difficulty of optimizing a single orthonormal orbital basis set from which CSF expansions would  describe efficiently all correlation effects  for a given physical state, and would produce reliable expectation values for any operator other than the Hamiltonian.  

To avoid these problems inherent to the use of a single common orthonormal orbital basis  for all correlation effects in the MCHF method, we propose to first optimize independent MCHF pair correlation functions (PCFs), bringing their own orthonormal one-electron basis. These PCFs are then coupled to each other through a low dimension generalized eigenvalue problem. A pioneer and inspiring work was done by Froese Fischer and Saxena~\cite{FroSax:74a,FroSax:75c} who introduced the separated-pair MCHF approach. The originality of the present study lies in the way the one-electron nonorthogonalities are treated for setting up the interaction matrix between the reference and the PCF spaces. For each coupling matrix element, we adopt the biorthogonal orbital transformations and CI eigenvectors counter-transformations, as originally proposed by Malmqvist~\cite{Mal:86a} and later extended to the spherical atomic symmetry by Olsen {\it et al.}~\cite{Olsetal:95a}.

The PCF interaction approach is tested, in the nonrelativistic approximation, on the ground state of beryllium, using a multireference description for the valence electrons. 	The results are compared with the complete active space (CAS) MCHF method.

\section{The variational method}

The MCHF method, that is extended and explored with respect to nonorthogonalites in the current work, 
can be derived from the variational principle.
A state $\Psi$ is an eigenstate of the Hamiltonian $H$ if and only if the energy functional
\begin{equation}
E[\Psi] \equiv \frac{\langle \Psi |H|\Psi \rangle}{\langle \Psi | \Psi \rangle}
\end{equation}
is left unchanged for any infinitesimal variation in the state at the point $\Psi$.
For the ground state or states that are the lowest of it's symmetry, the variational method gives a minimum
principle
\begin{equation}
E_0 \le \frac{\langle \Psi |H|\Psi \rangle}{\langle \Psi | \Psi \rangle},
\end{equation}
$E[\Psi]$ being the upper bound to the exact ground state energy $E_0$. Using a superposition ansatz
\begin{equation}
|\Psi \rangle = \sum_{i=1}^M c_i | \Phi_i \rangle
\end{equation}
for describing the model state in the subspace of the $M$ basis states $\Phi_i$, the variational parameters
$\{c_i\}$ can be determined by solving the generalized eigenvalue problem
\begin{equation}
{\bf Hc} = E{\bf Sc}
\label{GEP}
\end{equation}
with $H_{i,j} \equiv \langle \Phi_i|H| \Phi_j \rangle$ and
$S_{i,j} \equiv \langle \Phi_i|\Phi_j \rangle$
being, respectively, the Hamiltonian and overlap matrices.
The total energies are found as roots of the secular equation
\begin{equation}
\det({\bf H} - E {\bf S}) = 0.
\end{equation}
The use of the eigenfunctions $\{\Psi^{(k)}\}$ as a model subspace satisfying
\begin{equation}
\langle \Psi^{(k)}|\Psi^{(l)} \rangle = \delta_{k,l}~,~~~\langle \Psi^{(k)}|H|\Psi^{(l)} \rangle
= \epsilon_k \delta_{kl}
\end{equation}
is useful for the description of excited states since they
according to the Hylleraas-Undheim theorem \cite{HylUnd:30a,McD:33a} satisfy the
conditions
\begin{equation}
E_k \le \epsilon_k~,~~~\forall~k = 1,\ldots,M.
\end{equation}
That is, approximate eigenvalues obtained by diagonalizing the Hamiltonian
in a subspace can only be stabilized when the latter is enlarged.

\section{The nonrelativistic multiconfiguration Hartree-Fock method}
Starting from the nonrelativistic Hamiltonian for an $N$-electron system
\begin{equation}
H = \sum_{i=1}^N \left[ - \frac{1}{2} \nabla_i^2 - \frac{Z}{r_i} \right] + \sum_{i<j}^N \frac{1}{r_{ij}},
\end{equation}
the MCHF approach determines an approximate
wave function $\Psi$ for the state labeled $\gamma LS$ of the form
\begin{equation}
|\Psi(\gamma LS)\rangle = \sum_{i=1}^M c_i | \Phi(\gamma_i LS) \rangle,
\end{equation}
where $\gamma$ represents the dominant configuration and any additional quantum numbers required for
uniquely specifying the state being considered. The CSFs $\{\Phi(\gamma_j LS)\}$ 
are built from
a basis of one-electron spin-orbitals
\begin{equation}
\phi(nlm_lm_s) = \frac{1}{r} P(nl\,;\,r) Y_{lm_l}(\theta, \varphi)\chi_{m_s},
\end{equation}
where the radial distributions $\{P(nl\,;\,r)\}$ are to be determined. By applying the
variational principle one obtains a set of integro-differential MCHF equations
\begin{equation}
\left\{ \frac{d^2}{dr^2} + \frac{2}{r} \left[Z - Y(nl\,;\,r)\right]  - 
\frac{l(l+1)}{r^2} - \epsilon_{nl,nl} \right\} P(nl\,;\,r) \\
                   = \frac{2}{r} X(nl\,;\,r) + \sum_{n'\ne n} \epsilon_{nl,n'l} P(n'l\,;\,r)
\end{equation}
for the unknown radial distributions \cite{Froetal:97b}. The equations are coupled to each other
through the direct $Y$ and exchange $X$ potentials and the Lagrange multipliers $\epsilon_{nl,n'l}$. 
The Lagrange multipliers force the radial orbitals to be orthonormal  
within the same $l$ subspace. Under these conditions the configuration state functions are orthonormal
\begin{equation} 
\langle \Phi(\gamma_i LS)| \Phi(\gamma_j LS) \rangle = \delta_{i,j}.
\end{equation} 
The mixing coefficients $\{c_i\}$ appearing in the expansion over CSFs also enter in the form of the potentials
and can be determined by solving the CI problem 
\begin{equation}
{\bf Hc} = E {\bf c}
\end{equation}
for the current set of radial distributions. The MCHF and CI problems are solved iteratively until self-consistency
is reached for the radial distributions and for the selected CI-eigenvector.\medskip\\
The multiconfiguration method incorporates an extension allowing radial nonorthonormality,
which sometimes can be used to advantage. Evaluation of matrix elements when orbitals are nonorthogonal is complex, in general. In order to keep the energy expressions manageable a number of restrictions are imposed:
the orbitals within each CSF are mutually orthogonal, there are at most two subshells in $\Phi(\gamma_i LS)$ containing spectator electrons whose orbitals are nonorthogonal
to orbitals in the interacting function $\Phi(\gamma_j LS)$, if all the spectator electrons with nonorthogonal orbitals have the same $l$ value then there are at most two such electrons
in each of $\Phi(\gamma_i LS)$ and $\Phi(\gamma_j LS)$, the use of nonorthogonal orbitals is such that $\langle \Phi(\gamma_i LS)| \Phi(\gamma_j LS) \rangle = \delta_{i,j}$.
Here the term spectator refers to those electrons not directly involved in the interaction~\cite{Froetal:97b}.
Despite these restrictions, nonorthogonal orbitals have successfully been applied in a number of cases, describing major correlation effects
in a very compact way \cite{Godetal:98a,Godetal:97b}.

\section{Correlation and spatial location of orbitals}
\label{sec_spatial}

The MCHF method results from the application of the variational principle, and the solution depends strongly
on the energy functional or CSF expansion used to derive the MCHF equations. The direct influence of the CSF expansion on the shape
of the resulting radial orbital basis can be exploited to target specific effects, but could also bring undesirable
distortions of the wave function for the property of interest. To illustrate these effects let us look at the beryllium
$1s^2\,2s^2~^1S$ ground state. The valence pair-correlation function (PCF) is described
by considering all possible excitations of the valence electron pair
\begin{eqnarray}
\ket{\Lambda_{VV}}&=&\alpha_1\ket{1s^2\,2s^2\ ^1S} \nonumber \\ &&
+\sum_{n}\alpha_{n}\ket{1s^2\,2s\,ns\ ^1S}+\sum_{nl,n'l'}\alpha_{nl,n'l'}\ket{1s^2\,nl\,n'l'\ ^1S}.
\end{eqnarray}
The core-valence PCF is obtained by promoting one core electron together with one valence electron
\begin{eqnarray}
\ket{\Lambda_{CV}}&=&\beta_1\ket{1s^2\,2s^2\ ^1S} 
+\sum_{nl,n'l'}\beta_{nl,n'l'}\ket{1s\,2s\,nl\,n'l'\ ^1S}.
\end{eqnarray}
Finally, the core-core PCF takes into account the correlation within the core
\begin{eqnarray}
\ket{\Lambda_{CC}}&=&\gamma_1\ket{1s^2\,2s^2\ ^1S} \nonumber\\  &&
+\sum_{n}\gamma_{n}\ket{1s\,2s^2\,ns\ ^1S}+ \sum_{nl,n'l'}\gamma_{nl,n'l'}\ket{2s^2\,nl\,n'l'\ ^1S}.
\end{eqnarray}
Each of the above sums represent correlation between a particular pair of electrons, namely $(2s,2s)$ for valence correlation, $(1s,2s)$ for core-valence, and
$(1s,1s)$ for core-core correlation. For this reason they are called pair-correlation functions.
As far as terminology is concerned, pair-correlation   should be strictly used for double-excitations~\cite{LinSal:80a}. 
However, single-excitations are also included in our PCFs. Somewhat arbitrary we choose to include $\ket{1s^2\,2s\,ns\ ^1S}$ in the valence function 
and $\ket{1s\,2s^2\,ns \ ^1S} $ in the core-core function. Other choices are possible for the single-excitations. 
In the above case there is only one reference configuration, but applied to a multireference the PCFs capture 
major correlation effects in a very effective way~ \cite{Froetal:97b,MorFro:87a}. \medskip\\
The dependence of the orbitals on the type of correlation is illustrated in Figure 1. Here one clearly sees
the contraction of the correlation orbitals when going from a valence to a core-core correlation calculation. Unfortunately,
correlation effects are not additive, and interference occurs through multiple excitations. As we have described above one can easily 
target specific correlation effects through the CSF expansion. However, the resulting set of radial distributions 
is localized in such a way that it becomes inadequate for representing other types of correlation. In our four-electron system it
is for example not a meaningful to use core-core correlation orbitals for describing valence correlation and vice versa. 
The problem with the space localization of the correlation orbitals becomes more pronounced for larger systems with many electron subshells.\\ 

\begin{figure}[!h]
\begin{center}
\mbox{\includegraphics[scale=0.3]{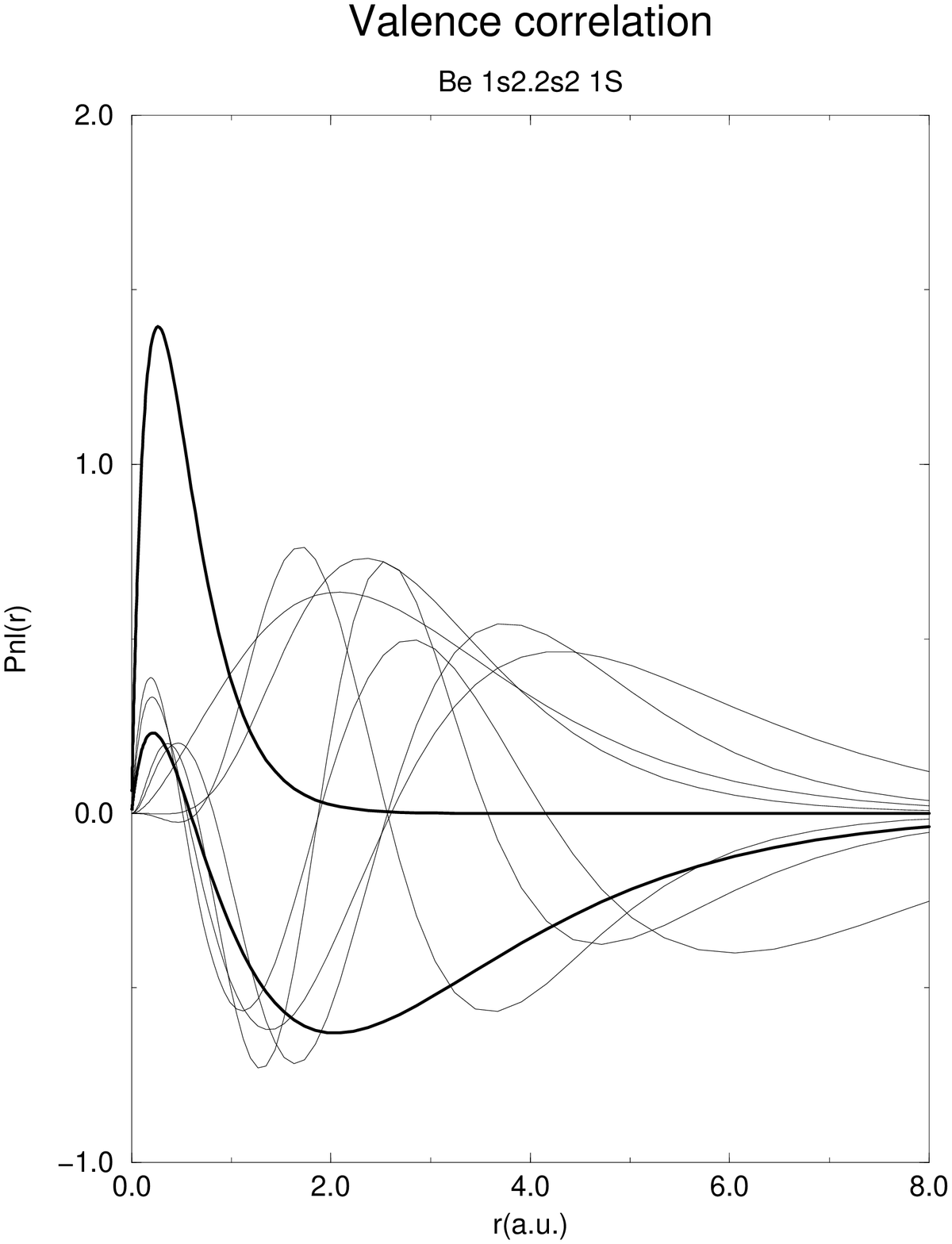}
\includegraphics[scale=0.3]{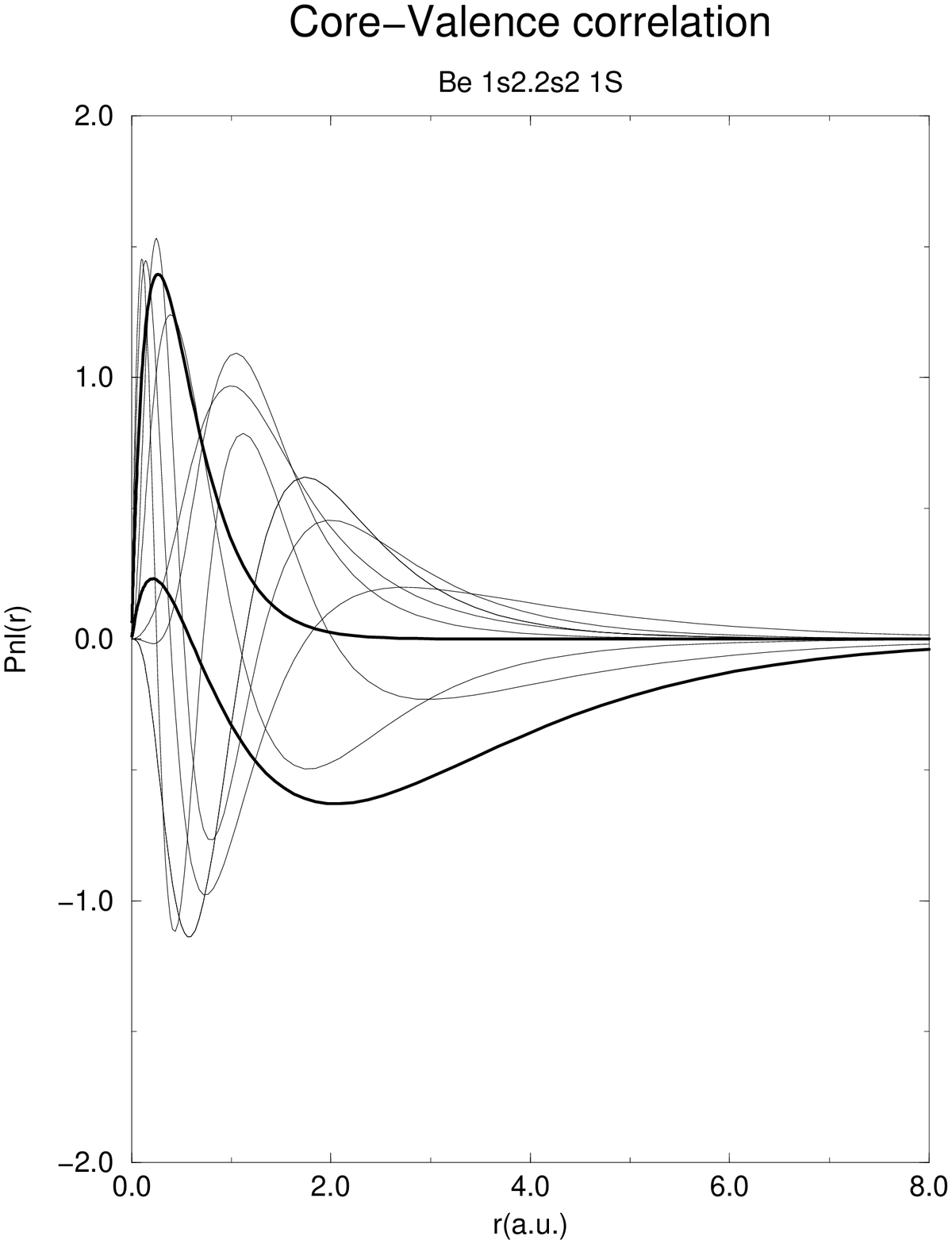}
\includegraphics[scale=0.3]{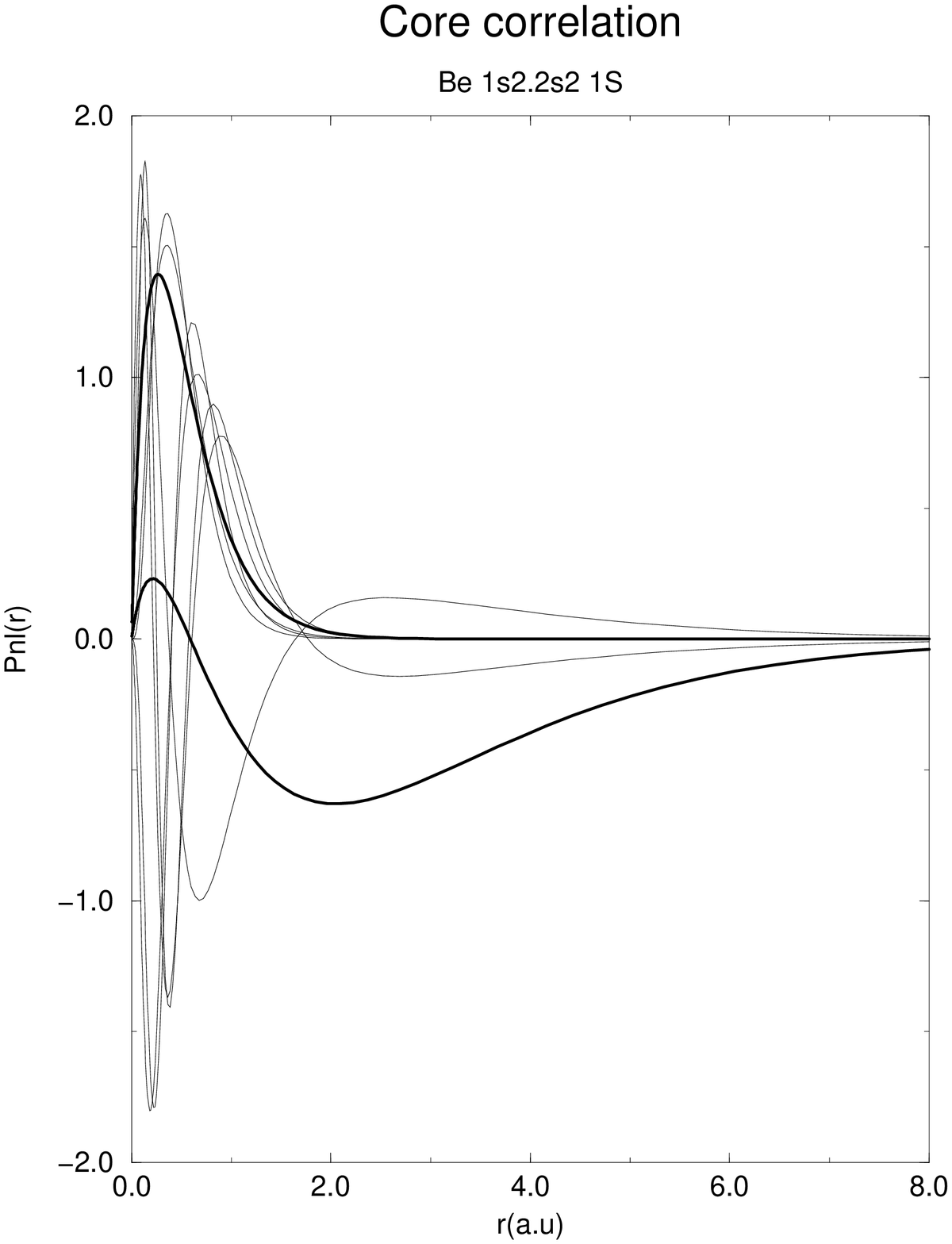}}
\caption{Contraction of the correlation orbitals when going from valence to core-valence and
core-core correlation MCHF calculations of Be $1s^2\,2s^2~^1S$. The two thick lines correspond
to the spectroscopic $1s$ (no node) and $2s$ (one node) orbitals. Other lines represents the radial distributions of the
correlation orbitals of the $n=4$ active set.}
\label{fig:delta_conv}
\end{center}
\end{figure}

\noindent 
Multiconfiguration expansions frequently target valence correlation and correlation between valence electrons and the outermost core electrons. Correlation within the core is neglected. Based on the combined valence and core-valence expansion a single orbital basis is determined. While this approach gives orthonormal configuration state functions, and allows standard methods to be used for the construction of radial matrix elements, it has some drawbacks. A large orbital basis is needed to saturate both the valence and core-valence correlation effects. This in turn leads to huge numbers of CSFs, many of which are unimportant. Another problem is uneven convergence patterns as new layers of orbitals, dependent on the gain in variational energy, may be more or less contracted, affecting computed properties differently. Finally, there are difficulties to effectively incorporate correlation within the core. A solution to these problems, taking the nonorthogonal extension of the MCHF package all the 
way, would be to represent the wave function as an expansion over CSFs
\begin{equation}
|\Psi \rangle = \sum_{i=1}^M c_i | \Phi_i^{opt} \rangle,
\end{equation}
where each CSF, depending on the correlation effect it describes, is built on an
orbital set optimized for this effect. Since different orbital sets are involved, the CSFs
are no longer orthonormal, and the expansion coefficients are obtained by solving the generalized eigenvalue problem.
The  Hamiltonian matrix elements coupling CSFs built on different and nonorthogonal orbital sets, could be evaluated using  their Slater determinant expansions and the L\"owdin's cofactor method \cite{Low:55a,Wee:92a}, as adopted in many atomic structure packages~\cite{Zat:96a,ZatFro:00b,Frietal:99a,IndDes:07a}. For CSF expansions, a nonorthogonal extension of the underlying Racah algebra~\cite{Fan:65a} is available~\cite{Hibetal:88a,HibFro:91a} but is not general enough. Moreover, both approaches quickly become prohibitive from a computational point of view.

\section{Pair-Correlation function interaction calculations}

A different way of handling nonorthogonalities is offered by the biorthonormal transformation technique
originally introduced by Malmqvist \cite{Mal:86a}. The transformation allows general matrix elements 
between PCFs to be computed fast and efficiently. To exploit this we represent the wave function, not as an expansion over CSFs, but as
an expansion over one or more reference CSFs $\ket{\Phi_{i}^r}$
and a number of known pair-correlation functions $\ket{\tilde{\Lambda}_j}$
built on separate and optimally localized orbital sets 
\begin{equation}
|\Psi \rangle =    
\sum_{i = 1}^g c_i \ket{\Phi_{i}^r} +  \sum_{j=1}^p  \tilde{c}_j\ket{\tilde{\Lambda}_j}. 
\end{equation}
To avoid redundancies in the representation, reference CSFs are discarded from the PCFs. In the
remaining sections we will use the notation $\ket{\Lambda_j}$ for PCFs including the reference CSFs and $\ket{\tilde{\Lambda}_j}$ for
PCFs where the reference CSFs have been discarded (weights of reference CSFs have been set to zero).
The Hamiltonian and overlap matrix elements between reference CSFs and PCFs and between
different PCFs are computed using the biorthonormal transformation. The expansion coefficients $\{c_i,\tilde{c}_j\}$ are obtained by
solving the corresponding generalized eigenvalue problem. The problem of computing a total wave function including 
valence, core-valence and core-core correlation
thus reduces to a series of separate pair-correlation problems
together with a relatively small generalized eigenvalue problem.
Practically, the calculations proceed in the following steps:
\begin{enumerate}
\item Perform an MCHF calculation for the reference expansion.
\item Keep the orbitals from the first step fixed and perform separate MCHF calculations for the different
PCFs including the reference.
\item Remove the reference CSFs from the PCFs by setting the weights to zero.
\item Loop over CSFs and PCFs, 
perform biorthogonal transformations and evaluate the Hamiltonian and overlap matrix elements.
\item Solve the generalized eigenvalue problem.
\end{enumerate}
This computational scheme has the advantage of using targeted and optimally localized orbital sets for each of the PCFs.
It is possible to include core correlation in a tractable way. Also, it is possible to determine the contribution to computed properties from
each of the PCFs. What is lost is some variational freedom in the coefficients when going from an expansion
of CSFs to and expansion over reference CSFs and PCFs. Below we describe the
biorthogonal transformation at heart of the method. 

\section{Biorthogonal transformations}

In the  above context, solving the generalized eigenvalue 
problem~(\ref{GEP}) of dimension~$(g+p)$ in the basis of the 
$g$~reference CSFs and $p$~PCFs
\begin{equation}
\{ \ket{\Phi_{1}^r} , \ket{\Phi_{2}^r}, \ldots \ket{\Phi_{g}^r} , \;
\ket{\tilde{\Lambda}_{1}} , \ket{\tilde{\Lambda}_{2}}, \ldots 
\ket{\tilde{\Lambda}_{p}}  \}
\end{equation}
requires the calculation of the Hamiltonian and Gram (overlap) matrix 
elements.
In the present approach, freezing the orbitals optimized for the MCHF 
reference function
guarantees the orthononormality of orbitals subsets involved in
$\langle \Phi_{i}^r | H | \tilde{\Lambda}_{j} \rangle $
and 
$\langle \Phi_{i}^r |\tilde{\Lambda}_{j} \rangle $
matrix elements, 
but one-electron nonorthogonalities definitely appear in off-diagonal 
matrix elements coupling different PCFs, i.e.
$ \langle \tilde{\Lambda}_{i} | H | \tilde{\Lambda}_{j} \rangle  $. 
In the most general case invoking other optimization strategies, 
one-electron nonorthogonalities will show up in all but the diagonal matrix 
elements. The biorthonormal approach is then applied for evaluating such 
matrix elements.

The biorthogonal transformation is explained in details in 
\cite{Olsetal:95a} focusing on the calculation of transition 
probabilities using nonorthogonal orbitals. This method has shown to be 
very efficient and useful, producing reliable atomic transition
data~\cite{FroTac:04a,Froetal:06a} through the atomic structure package 
ATSP2K~\cite{Froetal:07a}. It has also been implemented in 
GRASP2K~\cite{Jonetal:07a} for the calculation of transition amplitudes 
in the full relativistic context~\cite{Gra:07a}.
The idea is simple: two orbital sets that are not orthonormal to each other
are first transformed to become biorthonormal.
For a coupling matrix element $ \langle \tilde{\Lambda}_{l}  | H | 
\tilde{\Lambda}_{r} \rangle $ built in their own orbital basis
$ \{ \phi^L_i \}$ and $ \{ \phi^R_i \}$ that are not orthonormal
\begin{equation}
 \langle \phi^L_i \vert  \phi^R_j \rangle = S_{ij}^{LR} \; ,
\end{equation}
linear transformations\footnote{in our matrix notation, the orbitals $ 
\{ \phi_i \}$ are collected in row vectors $\mbox{\boldmath $ \phi $}$. }
\begin{equation}
\mbox{\boldmath $ \phi $}^A = \mbox{\boldmath $ \phi $}^L {\bf C}^{LA }
\hspace*{1cm} ; \hspace*{1cm}
\mbox{\boldmath $ \phi $}^B = \mbox{\boldmath $ \phi $}^R {\bf C}^{RB } \; ,
\end{equation}
are found to transform the two original orbital sets into two new 
biorthonormal sets
\begin{equation}
 \langle \phi^A_i \vert  \phi^B_j \rangle = \delta_{ij} \; .
 \label{biort}
\end{equation}
The advantage of the biorthonormality property (\ref{biort}) of the 
transformed orbital sets is that the evaluation of any matrix elements 
can proceed as in the orthonormal case, as originally found by Moshinsky 
and Seligman~\cite{MosSel:71a}.
There is an infinity of pairs of transformation matrices $({\bf C}^{LA 
},{\bf C}^{RB })$ that produce biorthonormal basis sets. In our approach 
\cite{Olsetal:95a}, the choice adopted is predicted by the restrictions 
on the configuration state function spaces used for 
$\tilde{\Lambda}_{l}$ and $\tilde{\Lambda}_{r}$.
We require the transformation matrices to be upper triangular, a 
suitable choice for estimating the effect of the orbital transformation 
on the mixing coefficients
$\{ \alpha^L_i \} \rightarrow \{ \alpha^A_i \} $ / $ \{ \alpha^R_i \} 
\rightarrow \{ \alpha^B_i \} $   giving the two representations of the 
$| \tilde{\Lambda}_{l} \rangle $ and $| \tilde{\Lambda}_{r} \rangle $ 
functions in both original and transformed (biorthonormal) basis sets
\begin{equation}
| \tilde{\Lambda}_{l} \rangle =
\sum_i \alpha^L_i    \ket{\Phi^L_{i}} = \sum_i \alpha^A_i 
\ket{\Phi^A_{i}} \; ,
\label{counter_L}
\end{equation}
\begin{equation}
| \tilde{\Lambda}_{r} \rangle =
\sum_i \alpha^R_i    \ket{\Phi^R_{i}} = \sum_i \alpha^B_i 
\ket{\Phi^B_{i}} \; .
\label{counter_R}
\end{equation}
Malmqvist~\cite{Mal:86a} has indeed shown that an upper-triangular 
orbital transformation matrix can be expressed as a finite sequence of 
single-orbital transformations, each expressing the new orbitals as a 
sum involving no higher-numbered orbital. Each such transformation step 
on the CI-expansion array is the same as the effect of a one-electron 
operator with de-excitations only.
 To avoid symmetry-breaking intermediates in the recursive 
transformation, some atomic symmetry refinement was made in 
\cite{Olsetal:95a} to express the transformation operator acting in the 
CSF space in the following suitable form
\begin{equation}
\left(
\sum_{N=0}^{2(2l+1)} \frac{1}{N \!} \; \hat{s}^N
\right) t_{nn}^{\hat{N}_n} \; ,
\end{equation}
with
\begin{equation}
\hat{s} =
\sum_{n' \neq n} \frac{t_{n'n}}{t_{nn}}
\left(
\sum_{m_l m_s}^{2(2l+1)}
a_{n'lm_lm_s}^{\dagger} \hat{a}_{nlm_lm_s}
\right)
\; .
\end{equation}
$\hat{N}_k$ is the occupation number of the $(nl)_k$-subshell and  ${\bf 
t}$ is the  matrix  defining the single orbital replacements sequence 
that can be calculated from the UL decomposition of $\left( {\bf S}^{LR} 
\right)^{-1}$. The knowledge of the action of the excitation operators 
in the  CSF spaces for both
$| \tilde{\Lambda}_{l} \rangle $ and $| \tilde{\Lambda}_{r} \rangle $
\begin{equation}
\left(
\sum_{m_l m_s}^{2(2l+1)}
a_{n'lm_lm_s}^{\dagger} \hat{a}_{nlm_lm_s} \right)
 \vert \Phi_i \rangle = \sum_{j} \;  A_{ij} \vert \Phi_j \rangle 
\end{equation}
is then enough to perform the countertransformations of the 
corresponding CI vectors. Starting from the coupled tensorial second 
quantized form of the  single-particle  Hamiltonian operator
\begin{equation}
H = - \sum_l \sqrt{2(2l+1)} \; \sum_{n',n} \left( a_{n'l}^{\dagger} 
a_{nl}  \right) ^{(00)}_{00} \: I_{n'l,nl},
\end{equation}
one  realizes \cite{Olsetal:95a} that $A_{ij} = - \sqrt{2(2l+1)}$ times 
the coefficient of the $ I_{n'l,nl}$ integral found in $H_{ij}$ (see also \cite{Borgetal_0907.2830v1}).

There is an important built-in constraint in the algorithm:  the wave 
function expansion spaces for both left and right should be ``closed 
under de-excitation'' to allow this class of transformation. 
Restricted active space~(RAS) and complete active space~(CAS) expansions 
~\cite{Olsetal:88a} satisfy this property, i.e. removing one electron 
from a subshell $nl$ and placing it in any subshell $n'l$ of the same 
spatial symmetry $(n' < n)$ generates a CSF that appears in the original 
configuration expansion. 
Other types of expansion can easily be complemented 
with CSFs to satisfy the closure condition.

\section{Code implementation}

The nonorthogonal pair-correlation method is implemented in a code module extending the ATSP2K package \cite{Froetal:07a}. The module contains 
routines for checking closure under de-excitation (\verb+lscud+), performing biorthogonal transformations (\verb+biotrans+), 
evaluating Hamiltonian and overlap matrix elements (\verb+biomatrix+), and solving the
generalized eigenvalue problem \cite{BIO-ATSP:09a}. Having determined the reference and the pair-correlation functions, removing the reference CSFs 
from the latter, the calculation proceeds as follows:
\begin{verbatim}
    loop over reference and pair-correlation functions
       call lscud
       call biotrans
       call biomatrix
    end loop
    call generalized eigenvalue solver
\end{verbatim}
To perform the biorthogonal transformation only one-electron coupling coefficients are needed, and thus this part is computationally fast.
The subsequent evaluation of the Hamiltonian and overlap matrix elements utilizes standard Racah algebra techniques in the biorthonormal basis. 
Since there are two orbital sets being biorthogonal, rather than just one set, normal symmetry relations for radial integrals do not hold and integration routines
need to be re-designed. 
The correctness of the implemented method was checked by comparing Hamiltonian and overlap matrix elements with the corresponding ones 
calculated using Slater determinant algebra in nonorthogonal bases \cite{Low:55a,Wee:92a}. Due to the complexity of the latter method
this could only be done for two-, three- and four-electron systems. The correctness of the codes can also be inferred from calculations using
artificially rotated orbital sets, where invariances in the representations are used. Tests for different types of systems showed that the
biorthonormal transformation is numerically robust, with negligible loss of accuracy for the matrix elements. 

\section{Exploring computational strategies}

The beryllium atom may be thought of as the ``benchmark'' atom. It is the first neutral system in which the three correlation effects, i.e. valence, core-valence and core correlation, appear and during the years much work has been devoted to understand and accurately describe these effects.
Although the goal of the present work is not to get the lowest absolute total energy of the beryllium ground state, it is appropriate to highlight  some important contributions with regard to chronology. Byron and Joachain \cite{ByrJoa:67a} decoupled the original four-electron problem into a series of helium-like equations describing pair correlation between electrons. A combined configuration-interaction-Hylleraas-type wave function study was performed by Sims and Hagstrom~\cite{SimHag:71a}. A rather complete correlation study of Be was presented by Froese Fischer and Saxena~\cite{FroSax:74a} who introduced the separated-pair MCHF approach (this latter work is undoubtedly the inspiration source of the present approach). Elaborate and highly accurate Be ground state total energies were predicted by Bunge~\cite{Bun:76a}. 	An iterative numeric procedure~\cite{Mar79a} was used to obtain pair functions applied to two-electron systems. Numerical many-body perturbation calculations on Be-like systems were 
	proposed by Salomonson {\it et al.}~\cite{Saletal:80a} adopting a multiconfigurational model space. A unified approach combining the multiconfiguration Hartree-Fock method and many-body perturbation 	theory was attempted by Morrison~\cite{Mor:86a,MorFro:87a,Mor:88b}.
	The coupled-cluster single- and double-excitation 	equations were solved numerically 
by Salomonson and \"Oster~\cite{SalOst:90a}. After the beryllium atom was ``revisited'' by a number of groups~\cite{Maretal:91a}, the nonrelativistic total energy of the Be ground state was estimated by Lindroth {\it et al.} \cite{Linetal:92a}. 
Accurate values for the nonrelativistic total energy of Be ground state have been obtained using the full-core plus correlation method \cite{Chuetal:93a}. A new correlation study by Froese Fischer~\cite{Fro:93a} was presented in the MCHF scheme, extending the study of $n$-expansion methods to four-electron systems. Energy levels and oscillator strengths were calculated by Weiss~\cite{Wei:95a} adopting a multireference superposition-of-configurations approach for describing core-correlation effects. Finite-element MCHF and GTO basis set expansions coupled with MRSDCI were used to estimate the beryllium electron affinity by Olsen {\it et al.}~\cite{Olsetal:94a}.
Although difficulties appear in computing matrix elements in the Hylleraas basis \cite{Kin:93a,Haretal:04a}, Hylleraas configuration-interaction calculations were performed on the nonrelativistic ground-state energy of the Be atom~\cite{BusKle:95a}.  
The most accurate energies for the ground state of the beryllium-like atoms have been obtained using the exponentially correlated Gaussian basis sets \cite{PacKom:04a,Staetal:09a}. The most recent work  \cite{Staetal:09a} reports the up-to-date lowest infinite mass nonrelativistic total energy of $-14.667~356~486~E_\textrm{h}$. \\

Below we explore different computational strategies for $1s^22s^2~^1S$ in beryllium based on separately optimized pair-correlation functions.
The results are compared with calculations utilizing a single orthonormal orbital set. Whereas many accurate computational schemes can only be applied
to small systems, the current method is directly generalizable to more complex systems and cases for which it currently is not possible to saturate the orbital basis (see section~\ref{sec_perspectives}). 

\subsection*{Monoreference $1s^22s^2~^1S$ - $(4\times 4)$ approach}
We start by generating the three separate PCFs: the valence, the core-valence, and the core-core   
\begin{eqnarray}
\ket{\Lambda_{VV}}&=&\alpha_1\ket{1s^2~2s^2\ ^1S} \nonumber \\ &&
+\sum_{n}\alpha_{n}\ket{1s^2~2s~ns\ ^1S}+\sum_{nl,n'l'}\alpha_{nl,n'l'}\ket{1s^2~nl~n'l'\ ^1S}\\
\ket{\Lambda_{CV}}&=&\beta_1\ket{1s^2~2s^2\ ^1S} \nonumber\\
&&+\sum_{nl,n'l'}\beta_{nl,n'l'}\ket{1s~2s~nl~n'l'\ ^1S}  \\
\ket{\Lambda_{CC}}&=&\gamma_1\ket{1s^2~2s^2\ ^1S} \nonumber\\ 
&&+\sum_{n}\gamma_{n}\ket{1s~2s^2~ns\ ^1S}+ \sum_{nl,n'l'}\gamma_{nl,n'l'}\ket{2s^2~nl~n'l'\ ^1S}
\end{eqnarray}
The $1s$ and $2s$ orbitals are taken from  an initial  HF calculation and kept frozen. All correlation orbitals are variational resulting
in three sets of orbitals. The energies and number of CSFs as function of the largest principal quantum number 
are displayed in \Tref{tab:VV_HF} for each of the PCFs.
The table shows, as expected, larger correlation energies for valence and core excitations, relatively to core-valence.

\begin{table} [!h]
\caption{\label{tab:VV_HF} Energies together with the number of CSFs for the valence (VV), 
core-valence (CV), and core-core (CC) PCFs. The ($\underline{1s}$, $\underline{2s}$) orbitals are from HF.
Correlation orbitals for each of the PCFs are determined in separate MCHF calculations.}
\begin{indented}
\item[]\begin{tabular}{@{}c | c r| c r | c r} \br
$n\le$ & E$_{VV}$ & \multicolumn{1}{c|}{$N_{CSF}$} & E$_{CV}$ & \multicolumn{1}{c|}{$N_{CSF}$} & E$_{CC}$ & \multicolumn{1}{c}{$N_{CSF}$} \\ \hline
HF  & $-14.573~023~17$ & $1$   & $-14.573~023~17$ &  $1$   & $-14.573~023~17$ & $1$   \\
$2$ & $-14.616~062~66$ & $2$   & $-14.574~237~22$ &  $2$   & $-14.595~158~32$ & $2$   \\
$3$ & $-14.618~619~53$ & $7$   & $-14.578~062~34$ &  $6$   & $-14.612~266~39$ & $7$   \\
$4$ & $-14.618~990~08$ & $16$  & $-14.578~781~57$ &  $19$  & $-14.614~398~07$ & $16$  \\
$5$ & $-14.619~083~21$ & $30$  & $-14.578~960~23$ &  $40$  & $-14.615~041~50$ & $30$  \\
$6$ & $-14.619~121~18$ & $50$  & $-14.579~021~81$ &  $72$  & $-14.615~315~90$ & $50$  \\
$7$ & $-14.619~138~74$ & $77$  & $-14.579~050~02$ &  $117$ & $-14.615~455~15$ & $77$  \\
$8$ & $-14.619~148~03$ & $112$ & $-14.579~064~66$ &  $177$ & $-14.615~530~36$ & $112$ \\
$9$ & $-14.619~153~81$ & $156$ & $-14.579~071~26$ &  $254$ & $-14.615~580~19$ & $156$ \\
$10$& $-14.619~157~24$ & $210$ & $-14.579~076~87$ &  $350$ & $-14.615~611~88$ & $210$ \\
\br
\end{tabular}
\end{indented}
\end{table}
Denoting the PCFs in which the reference CSF $|1s^2\,2s^2\ ^1S \rangle$ has been removed by setting the corresponding expansion coefficient to zero by,
respectively, $\ket{\tilde{\Lambda}_{VV}}$, $\ket{\tilde{\Lambda}_{CV}}$, and $\ket{\tilde{\Lambda}_{CC}}$, we write the wave function as
\begin{equation}
|\Psi \rangle = c_{1s^22s^2} \ket{1s^2~2s^2\ ^1S} + c_{VV} \ket{\tilde{\Lambda}_{VV}} + c_{CV}\ket{\tilde{\Lambda}_{CV}} + c_{CC}\ket{\tilde{\Lambda}_{CC}}.
\end{equation}
The expansion coefficients and the total energies are obtained 
by solving the corresponding $4 \times 4$ generalized eigenvalue problem, see 
\Tref{tab:BIO_HF_4x4_2}.  
The total energies are compared with the SD-MCHF results obtained from the ``traditional'' approach, i.e. - i) generating the CSF expansion by single- and double-excitations from the reference state, and - ii) varying a common set of the orbitals representing the three correlation effects. 
The convergence with respect to the increasing orbital set
is initially much faster using the nonorthogonal pair-correlation approach, but as the orbital basis start to saturate the different correlation effects, the traditional 
method gives a lower total energy. Looking at the results one should bear in mind that, for a given active set specified by the largest principal
quantum number $n$, the number of variational orbitals is larger by a factor three in the pair-correlation approach, while the size of the PCF expansions  
are smaller than the traditional SD-MCHF expansion by roughly the same factor.

\begin{table}
\caption{\label{tab:BIO_HF_4x4_2} Solution of the $(4 \times 4)$ generalized eigenvalue problem.
The energies are compared with the SD-MCHF results based on a single orthonormal orbital set.}
\begin{indented}
\item[]\begin{tabular}{@{}c | c  c c c | c c} \br
	$n \leq $ & $c_{1s^22s^2}$ & $c_{VV}$ & $c_{CV}$ & $c_{CC}$ & $E_{4\times4}$  & $E_{SD-MCHF}$ \\
\hline
$2$  & $0.955006$ & $0.294901$& $1.1855[-2]$ & $2.9271[-2]$ & $-14.636~852~03$ & $-14.616~852~26$\\
$3$  & $0.959405$ & $0.278621$& $2.1230[-2]$ & $3.8235[-2]$ & $-14.658~281~46$ & $-14.651~174~28$\\
$4$  & $0.959708$ & $0.277380$& $2.2430[-2]$ & $3.8948[-2]$ & $-14.661~298~22$ & $-14.657~844~72$\\
$5$  & $0.959832$ & $0.276919$& $2.2651[-2]$ & $3.9046[-2]$ & $-14.662~171~17$ & $-14.660~930~03$\\
$6$  & $0.959920$ & $0.276607$& $2.2714[-2]$ & $3.9061[-2]$ & $-14.662~527~85$ & $-14.661~861~16$\\
$7$  & $0.959978$ & $0.276406$& $2.2737[-2]$ & $3.9062[-2]$ & $-14.662~704~55$ & $-14.662~413~74$\\
$8$  & $0.960006$ & $0.276305$& $2.2746[-2]$ & $3.9059[-2]$ & $-14.662~799~45$ & $-14.662~742~11$\\
$9$  & $0.960024$ & $0.276246$& $2.2750[-2]$ & $3.9057[-2]$ & $-14.662~858~90$ & $-14.662~908~34$\\
$10$ & $0.960035$ & $0.276206$& $2.2752[-2]$ & $3.9055[-2]$ & $-14.662~897~85$ & $-14.663~013~15$\\
\br                                                                
\end{tabular}
\end{indented}
\end{table}

\subsection*{Multireference $1s^2\; \{2s ,  2p\}^2~^1S$ - $(5\times 5)$ approach}
\label{section_MR_(2s,2p)2_4x4}
In this section we investigate a multireference calculation based on the Layzer's complex $1s^2\; \{2s^2 +  2p^2 \}~^1S$. The CSF list is generated by making simple and double excitations on this complex to an active set of orbitals. The CSFs are then arranged into the three lists, one for each correlation type,
\begin{eqnarray}
\ket{\Lambda_{VV}}&=&\alpha_1\ket{1s^2~2s^2\ ^1S}+\alpha_2\ket{1s^2~2p^2\; ^1S} \nonumber \\ &&
+\sum_{n}\alpha_{n}\ket{1s^2~2s~ns\ ^1S}+\sum_{m}\alpha_{m}\ket{1s^2~2p~mp\ ^1S} \nonumber \\ &&
+\sum_{nl,n'l'}\alpha_{nl,n'l'}\ket{1s^2~nl~n'l'\ ^1S}\\
\ket{\Lambda_{CV}}&=&\beta_1\ket{1s^2~2s^2\ ^1S}+\beta_2\ket{1s^2~2p^2\; ^1S}\nonumber\\
&&+\sum_{nl,n'l'}\beta_{nl,n'l'}\ket{1s~2s~nl~n'l'\ ^1S}\nonumber\\
&&+\sum_{nl,n'l'}\beta'_{nl,n'l'}\ket{1s~2p~nl~n'l'\ ^1S}\\
\ket{\Lambda_{CC}}&=&\gamma_1\ket{1s^2~2s^2\ ^1S}+\gamma_2\ket{1s^2~2p^2\; ^1S}\nonumber\\
&&+\sum_{n}\gamma_{n}\ket{1s~2s^2~ns\ ^1S}+\sum_{m}\gamma_{m}\ket{1s~2p^2~mp\ ^1S} \nonumber \\ 
&&+\sum_{nl,n'l'}\gamma_{nl,n'l'}\ket{2s^2~nl~n'l'\ ^1S}+\sum_{nl,n'l'}\gamma'_{nl,n'l'}\ket{2p^2~nl~n'l'\ ^1S}.
\end{eqnarray}
This kind of formal writing shows us that the CSF list corresponding to the valence correlation is rigorously the same as the one in the single reference case. However the content of the two other lists change, they now include simple, double, triple and quadruple excitations in comparison with the fundamental configuration $1s^2 2s^2~^1S$. 
Table \ref{tab:VV_HF_mref{1s,2s,2p}} contains the results of the three independent pair-correlation MCHF calculations in which the $1s$, $2s$, $2p$ orbitals are frozen from the
calculation of the multireference. The remaining correlation orbitals are completely variational in all the calculations.

\begin{table} [!h]
\caption{\label{tab:VV_HF_mref{1s,2s,2p}} Energies together with the number of CSFs for the multireference $1s^2\; \{2s ,  2p\}^2~^1S$ valence (VV), 
core-valence (CV), and core-core (CC) PCFs. The $\underline{1s}$, $\underline{2s}$, $\underline{2p}$ orbitals are 
from the multireference MCHF calculation. Correlation orbitals (except $2p$) for each of the PCFs are determined in separate MCHF calculations.}
\begin{indented}
\item[]\begin{tabular}{@{}c | c r |c r | c r} \br
$n\le$ & E$_{VV}$ & \multicolumn{1}{c|}{$N_{CSF}$} & E$_{CV}$ & \multicolumn{1}{c|}{$N_{CSF}$} & E$_{CC}$ & \multicolumn{1}{c}{$N_{CSF}$} \\ 
\hline
MR  & $-14.616~845~32$ &  2    & $-14.616~845~32$ &  2    & $-14.616~845~32$ &  2      \\
$3$ & $-14.618~905~91$ & $7$   & $-14.619~426~25$ & $12$  & $-14.654~311~44$ & $22$    \\
$4$ & $-14.619~124~07$ & $16$  & $-14.621~265~76$ & $40$  & $-14.658~148~75$ & $58$   \\
$5$ & $-14.619~200~57$ & $30$  & $-14.621~676~31$ & $93$  & $-14.659~044~75$ & $119$   \\
$6$ & $-14.619~234~08$ & $50$  & $-14.621~766~58$ & $177$ & $-14.659~373~26$ & $211$   \\
$7$ & $-14.619~251~00$ & $77$  & $-14.621~801~87$ & $298$ & $-14.659~517~62$ & $340$   \\
$8$ & $-14.619~260~03$ & $112$ & $-14.621~819~73$ & $462$ & $-14.659~609~71$ & $512$   \\
$9$ & $-14.619~265~73$ & $156$ & $-14.621~828~19$ & $675$ & $-14.659~662~96$ & $733$   \\
$10$& $-14.619~268~69$ & $210$ & $-14.621~833~99$ & $943$ & $-14.659~696~30$ & $1009$  \\
\br                             
\end{tabular}
\end{indented}
\end{table}

Again, using the notation
$\ket{\tilde{\Lambda}_{VV}}$, $\ket{\tilde{\Lambda}_{CV}}$, and $\ket{\tilde{\Lambda}_{CC}}$ for the PCFs where the weights of the reference CSFs have been set to zero,
the wave function is written
\begin{eqnarray}
|\Psi \rangle &=& c_{1s^22s^2} \ket{1s^2~2s^2\ ^1S} + c_{1s^22p^2} \ket{1s^2~2p^2\ ^1S} \nonumber \\ && 
+ c_{VV} \ket{\tilde{\Lambda}_{VV}} + c_{CV}\ket{\tilde{\Lambda}_{CV}} + c_{CC}\ket{\tilde{\Lambda}_{CC}}.
\end{eqnarray}
The expansion coefficients and the total energies are obtained 
by solving the  $5 \times 5$ generalized eigenvalue problem. The coefficients and the energies are reported in Table~\ref{tab:BIO_HF_mref{1s,2s,2p}_5X5} as functions of the largest
principal quantum number in the expansions. The energies are compared with values from traditional SD-MR-MCHF and CAS-MCHF calculations. In the $4 \times 4$
approach $|1s^2\,2p^2\ ^1S\rangle$ entered the $\ket{\tilde{\Lambda}_{VV}}$ valence PCF, which obtained a comparatively large weight.
Separating out $|1s^2\,2p^2\ ^1S\rangle$ increases the variational freedom through the expansion coefficient $c_{1s^22p^2}$ and the valence PCF becomes
just a small correction. Looking at the energies we see a dramatic improvement. The present  $5 \times 5$ energy for $n=10$ is
comparable to the $n=10$ CAS-MCHF energy based on an expansion of more than 650~000 CSFs. The effort for the CAS-MCHF calculation is huge, and the case
has to run for days on a cluster. In contrast, generating the pair-correlation functions,
constructing the Hamiltonian and overlap matrices, and solving the $5 \times 5$ generalized eigenvalue problem is very fast and is easily done on a PC.  

\begin{table}[!h]
\caption{\label{tab:BIO_HF_mref{1s,2s,2p}_5X5} Solution of the $(5 \times 5)$ generalized eigenvalue problem.
The energies are compared with SD-MR-MCHF and CAS-MCHF results based on a single orthonormal orbital set.}
\begin{indented}
\item[]\begin{tabular}{@{}c | c  c c c c c c c c || c} \br
   $n \leq $ &$c_{1s^22s^2}$ & $c_{1s^22p^2}$ & $c_{VV}$&$c_{CV}$&$c_{CC}$  \\
\hline
$3$ & $0.952146$ & $0.299633$ & $4.2579 [-2]$& $1.7789 [-2]$ & $3.8845 [-2]$  \\
$4$ & $0.952674$ & $0.297365$ & $4.3498 [-2]$& $2.1246 [-2]$ & $4.0538 [-2]$  \\
$5$ & $0.952773$ & $0.296950$ & $4.3639 [-2]$& $2.1823 [-2]$ & $4.0768 [-2]$  \\
$6$ & $0.952832$ & $0.296749$ & $4.3642 [-2]$& $2.1922 [-2]$ & $4.0805 [-2]$  \\
$7$ & $0.952864$ & $0.296644$ & $4.3629 [-2]$& $2.1956 [-2]$ & $4.0817 [-2]$  \\
$8$ & $0.952885$ & $0.296577$ & $4.3615 [-2]$& $2.1968 [-2]$ & $4.0816 [-2]$  \\
$9$ & $0.952897$ & $0.296543$ & $4.3603 [-2]$& $2.1972 [-2]$ & $4.0814 [-2]$  \\
$10$ &$0.952902$ & $0.296527$ & $4.3597 [-2]$& $2.1973 [-2]$ & $4.0813 [-2]$  \\
\hline
$n \leq $ &  $E_{5 \times 5}$ &  $E_{MR-SD-MCHF}$  & $E_{CAS-MCHF}$ \\
\hline
$3$ &  $-14.658~887~01$ & $-14.654~399~79$ & $-14.654~414~59$ \\
$4$ &  $-14.664~774~35$ & $-14.661~865~68$ & $-14.661~403~17$ \\
$5$ &  $-14.666~173~94$ & $-14.664~721~18$ & $-14.664~839~93$ \\
$6$ &  $-14.666~628~18$ & $-14.665~938~59$ & $-14.666~067~32$ \\
$7$ &  $-14.666~826~48$ & $-14.666~407~90$ & $-14.666~541~14$ \\
$8$ &  $-14.666~945~78$ & $-14.666~722~13$ & $-14.666~857~41$ \\
$9$ &  $-14.667~013~54$ & $-14.666~876~64$ & $-14.667~012~75$ \\
$10$ & $-14.667~056~14$ & $-14.666~975~55$ & $-14.667~114~52$  \\
\br            
\end{tabular}  
\end{indented}
\end{table}

\subsection*{Multireference $1s^2\;  \{2s, 2p, 3s, 3p, 3d \}^2~^1S$ - $(8 \times 8)$ approach}
\label{section_MR_(n=3)_8x8}
Here we extended the multireference to the $n=3$ complex $1s^2\; \{2s^2 +  2p^2 + 3s^2 + 3p^2 + 3d^2\}~^1S$. As in the previous section we treat each component of the multireference
as a subspace of the interaction matrix, increasing the dimension of the generalized eigenvalue problem to 8. The results of the separate pair-correlation MCHF 
calculations are displayed in Table~\ref{tab:VV_HF_mref_RR}.  Even if the size of the largest expansion has increased, it is still a very small
problem. Comparing with Table~\ref{tab:VV_HF_mref{1s,2s,2p}}, we see that it is the core-valence and core correlation PCFs that now give lower energies.

\begin{table} [!h]
\caption{\label{tab:VV_HF_mref_RR} Energies together with the number of CSFs for the multireference $1s^2\; \{2s ,2p, 3s, 3p, 3d\}^2~^1S$ valence (VV), 
core-valence (CV), and core-core (CC) PCFs. The $\underline{nl}$, ($n \le 3$) orbitals are 
from the multireference MCHF calculation. Remaining correlation orbitals for each of the PCFs are determined in separate MCHF calculations.}
\begin{indented}
\item[]\begin{tabular}{@{}c | c r| c r | c r} \br
$n\le$ & E$_{VV}$ & \multicolumn{1}{c|}{$N_{CSF}$} & E$_{CV}$ & \multicolumn{1}{c|}{$N_{CSF}$} & E$_{CC}$ & \multicolumn{1}{c}{$N_{CSF}$} \\ \hline
MR  & $-14.618~914~67$ & $5$   & $-14.618~914~67$ & $5$    & $-14.618~914~67$ & $5$\\
$4$ & $-14.619~126~28$ & $16$  & $-14.622~479~34$ & $70$   & $-14.656~953~43$ & $153$    \\
$5$ & $-14.619~201~47$ & $30$  & $-14.623~667~44$ & $192$  & $-14.660~544~82$ & $352$    \\
$6$ & $-14.619~234~70$ & $50$  & $-14.623~918~77$ & $403$  & $-14.661~285~99$ & $667$    \\
$7$ & $-14.619~251~52$ & $77$  & $-14.623~977~10$ & $721$  & $-14.661~532~51$ & $1122$   \\
$8$ & $-14.619~260~43$ & $112$ & $-14.623~999~33$ & $1164$ & $-14.661~642~87$ & $1743$   \\
$9$ & $-14.619~266~16$ & $156$ & $-14.624~009~07$ & $1750$ & $-14.661~702~00$ & $2555$   \\
$10$& $-14.619~269~59$ & $210$ & $-14.624~015~40$ & $2497$ & $-14.661~737~42$ & $3583$   \\
\br           
\end{tabular} 
\end{indented}
\end{table}

\begin{figure}
\begin{center}
\includegraphics[scale=0.6]{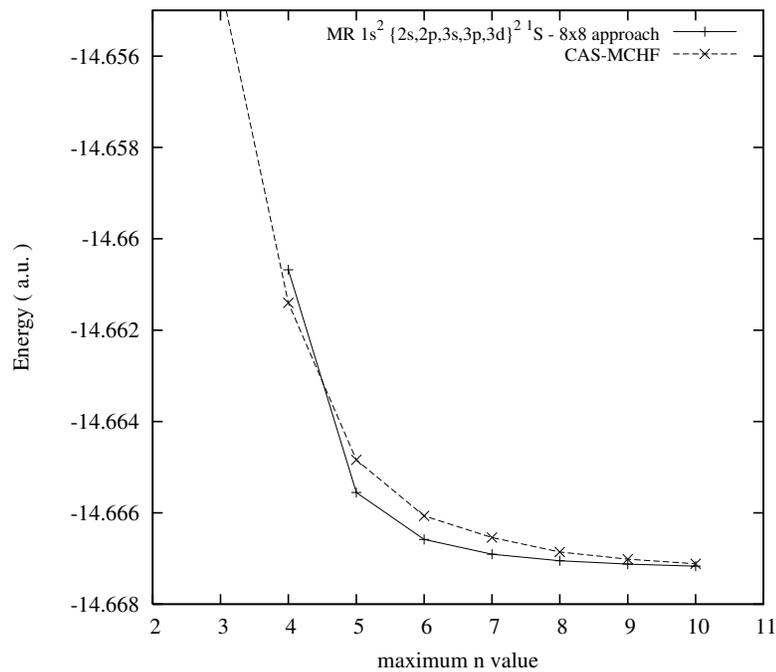}
\caption{Energies of the $1s^2\; \{2s ,2p, 3s, 3p, 3d\}^2~^1S$ multireference plus VV, CV, CC pair-correlation 
and the CAS-MCHF expansions as functions the principal quantum number $n$.}
\label{fig:Mref_3d}
\end{center}
\end{figure}

In this case the wave function has the form
\begin{eqnarray}
|\Psi \rangle &=& \sum_{nl, n \le 3} c_{1s^2nl^2} \ket{1s^2~nl^2\ ^1S} \nonumber \\ && 
+ c_{VV} \ket{\tilde{\Lambda}_{VV}} + c_{CV}\ket{\tilde{\Lambda}_{CV}} + c_{CC}\ket{\tilde{\Lambda}_{CC}}.
\end{eqnarray}
The expansion coefficients and the total energies are obtained 
by solving the  $8 \times 8$ generalized eigenvalue problem. The coefficients and the energies are reported in Table~\ref{tab:BIO_HF_mref_RR_5x5} as functions of the largest
principal quantum number $n$ in the expansions. The energies are compared with values from traditional CAS-MCHF calculations, which give the lowest possible
energies that can be obtained from a single orthogonal orbital set. Even for $n=10$, that is a basis consisting of 53 orbitals in the orthogonal case, the
pair-correlation approach gives a lower total energy. The convergence with respect to the largest principal quantum number $n$ is graphically displayed
in figure~\ref{fig:Mref_3d}. Here one clearly sees the slow saturation rate of the energy for the CAS-MCHF calculation. For larger systems with more subshells,
the difference in saturation rate between calculations using nonorthogonal PCFs and traditional calculations built on a single orbital set will be much greater. 
Note that although the PCF interaction total energy is lower than the CAS-MCHF one, it is still $2.10^{-4} E_{\textrm{h}}$ above the Stanke {\it et al.}'s result~\cite{Staetal:09a}. Remembering that both the PCF- and CAS-MCHF expansions are $l$-truncated ($l < 10$), this is expected due to the slow convergence rate~($E_l - E_{l-1} \simeq \mathcal{O} (l+1/2)^{-4}$) with respect to $l$ ~\cite{Hil:85a}.

\begin{table}[!h]
\caption{\label{tab:BIO_HF_mref_RR_5x5} Solution of the $(8 \times 8)$ generalized eigenvalue problem.
The energies are compared with CAS-MCHF results based on a single orthonormal orbital set.}
\begin{indented}
\item[]\begin{tabular}{@{}c | c  c c c c c c c c || c} \br
   $n \leq $ &$c_{1s^22s^2}$ & $c_{1s^22p^2}$ & $c_{1s^23s^2}$ & $c_{1s^23p^2}$ & $c_{1s^23d^2}$ \\
\hline
$4$ &  $0.953103$ & $0.295966$ & $-4.0663 [-2]$ & $5.8264 [-3]$ & $-1.7149 [-2]$ \\
$5$ &  $0.953210$ &  $0.295275$ & $-4.0808 [-2]$ & $5.1629 [-3]$ & $-1.6996 [-2]$  \\
$6$ &  $0.953281$ &  $0.295006$ & $-4.0807 [-2]$ & $5.1175 [-3]$ & $-1.6869 [-2]$  \\
$7$ &  $0.953322$ &  $0.294869$ & $-4.0783 [-2]$ & $5.0807 [-3]$ & $-1.6819 [-2]$  \\
$8$ &  $0.953342$ & $0.294804$ & $-4.0769 [-2]$ & $5.0626 [-3]$ & $-1.6789 [-2]$  \\
$9$ &  $0.953353$ & $0.294771$ & $-4.0763 [-2]$ & $5.0516 [-3]$ & $-1.6772 [-2]$  \\
$10$ & $0.953362$ & $0.294743$ & $-4.0758 [-2]$ & $5.0409 [-3]$ & $-1.6759 [-2]$  \\
\hline
$n \leq $ & $c_{VV}$&$c_{CV}$&$c_{CC}$ & $E_{8 \times 8}$ &  $E_{CAS-MCHF}$ \\
\hline
$4$ &   $8.5613 [-3]$& $2.0715 [-2]$ & $3.8931 [-2]$ & $-14.660~679~48$ & $-14.661~403~17$\\
$5$ &   $9.6715 [-3]$& $2.1911 [-2]$ & $4.0595 [-2]$ & $-14.665~553~46$ & $-14.664~839~93$\\
$6$ &   $9.8881 [-3]$& $2.2103 [-2]$ & $4.0808 [-2]$ & $-14.666~582~83$ & $-14.666~067~32$\\
$7$ &   $9.9702 [-3]$& $2.2135 [-2]$ & $4.0844 [-2]$ & $-14.666~905~87$ & $-14.666~541~14$\\
$8$ &   $9.9847 [-3]$& $2.2148 [-2]$ & $4.0849 [-2]$ & $-14.667~047~86$ & $-14.666~857~41$\\
$9$ &   $9.9822 [-3]$& $2.2151 [-2]$ & $4.0848 [-2]$ & $-14.667~122~76$ & $-14.667~012~75$\\
$10$&   $9.9793 [-3]$& $2.2151 [-2]$ & $4.0846 [-2]$ & $-14.667~168~08$ & $-14.667~114~52$  \\
\br
\end{tabular}
\end{indented}
\end{table} 

\clearpage

\section{Conclusion and perspectives} \label{sec_perspectives}
Variational calculations based on ``explicitly correlated Gaussians'' are quite impressive but unfortunately scale as $N!$, where $N$ is the number of electrons. As mentioned by Stanke {\it et al.}~\cite{Staetal:09a}, calculations
with fully correlated basis functions would require huge amounts of computer time for systems with more than six electrons. For the latter, the numerical MCHF method is widely used, generally limited to the optimization of a single orthonormal set of orbitals. 
However, the relaxation of one-electron orbitals and orthogonality constraints has been shown to be beneficial in many applications \cite{Fro:73a,Vaeetal:92a,Godetal:97b,ZatFro:00a}. 
The present work demonstrates the effectiveness of relaxing one-electron orthogonality constraints in the MCHF orbital optimization of different PCFs, yielding a new
method for treating correlation.
Applied to the ground state of beryllium the new 
method gives total energies that are
lower than the ones from traditional CAS-MCHF calculations using large single orbital sets with principal 
quantum numbers up to 10. Whereas many accurate computational schemes can only be applied
to small systems, the current method is directly generalizable to more complex systems 
and cases for which it currently is not possible to saturate a single orbital basis for describing different types of correlation contributing to the total energy, or different type of operators. 
It is fair to say that we now have the possibility to account for, in a balanced way, correlation deep down in the atomic core in variational calculations.

In this study we have only looked at a few aspects of separately optimized PCFs. Following Froese Fischer and Saxena \cite{FroSax:74a,FroSax:75c} 
one could further refine the method and define PCFs for each two-electron coupling. One would then, for example, differentiate between
\begin{equation}
\ket{\Lambda_{CV}} = \alpha_1\ket{1s^2\,2s^2\ ^1S} +\sum_{nl,n'l'}\alpha_{nl,n'l'}\ket{1s\,2s\ ^1S\,(nl\,n'l'\ ^1S)\ ^1S}
\end{equation}
and
\begin{equation}
\ket{\Lambda_{CV}} = \beta_1\ket{1s^2\,2s^2\ ^1S} +\sum_{nl,n'l'}\beta_{nl,n'l'}\ket{1s\,2s\ ^3S\,(nl\,n'l'\ ^3S)\ ^1S}.
\end{equation}
It would also be possible to introduce some single-electron excitation functions for describing core-opening effects crucial for hyperfine structure and
other one-electron operator quantities. The method with separately optimized pair-correlation functions
can be extended to a spectrum, describing several states at the same time. 
The atomic state function is then given by the superposition 
\begin{equation}
|\Psi \rangle =  \sum_{r=1}^{M} \left( 
\sum_{i = 1}^{g_r} c^r_i\ket{\Phi_i^r} +  \sum_{j=1}^{p_r}  \tilde{c}^r_j \ket{\tilde{\Lambda}_{j}^r} \right).
\end{equation} 
Energies and expansion coefficients of the states are obtained by selected solutions of the eigenvalue 
problem.

A lot of effort has been put \cite{BIO-ATSP:09a} in the general adaptation of ATSP2K \cite{Froetal:07a} for allowing separately optimized PCFs, and it is now possible to compute the expectation of any one- and two-electron operator, including the Breit-Pauli corrections, in the new wave function representation. The biorthogonal transformation for handling nonorthogonalities is applicable also in the fully relativistic case. This method with separately optimized PCFs is currently implemented \cite{BIO-ATSP:09a} in GRAPS2K~\cite{Jonetal:07a}.

\ack

S. Verdebout has a F.R.I.A. fellowship from the F.R.S.-FNRS Fund for Scientific Research.
M. Godefroid thanks the Communaut\'e fran\c{c}aise of Belgium (Action de Recherche Concert\'ee)
and the Belgian National Fund for Scientific Research (FRFC/IISN Convention) for financial support.

\clearpage
%
%

\section*{References}

\end{document}